# Robust superconductivity with large upper critical field in $Nb_2PdS_5$

Rajveer Jha, Brajesh Tiwari, Poonam Rani, Hari Kishan, and V.P.S. Awana[*]

National Physical Laboratory (CSIR) Dr. K. S. Krishnan Road, New Delhi-110012, India

**Abstract**

We report synthesis, structural details and complete superconducting characterization of very recently discovered [1] $Nb_2PdS_5$ new superconductor. The synthesized compound is crystallized in monoclinic structure with *C*2/*m* (#12) space group. Bulk superconductivity is seen in both *ac/dc* magnetic susceptibility and electrical resistivity measurements with superconducting transition temperature ($T_c$) at 6K. The upper critical field ($H_{c2}$), being estimated from high field magneto-transport [$\rho(T)H$] measurements is above 240kOe. The estimated $H_{c2}(0)$ is clearly above the Pauli paramagnetic limit of ~$1.84T_c$. Heat capacity ($C_p$) measurements show clear transition with well defined peak at $T_c$, but with lower jump than as expected for a *BCS* type superconductor. The Sommerfield constant ($\gamma$) and Debye temperature ($\Theta_D$) as determined from low temperature fitting of $C_P(T)$ data are 32mJ/mole-K$^2$ and 263K respectively. Hall coefficients and resistivity in conjugation with electronic heat capacity indicates multiple gap superconductivity signatures in $Nb_2PdS_5$. We also studied the impact of hydrostatic pressure (0–1.97Gpa) on superconductivity of $Nb_2PdS_5$ and found nearly no change in $T_c$ for the given pressure range.

[*]**Corresponding Author**
Dr. V. P. S. Awana, Senior Scientist
E-mail: awana@mail.npindia.org
Ph. +91-11-45609357, Fax-+91-11-45609310
Homepage www.freewebs.com/vpsawana/

**Introduction:**

Recent discovery of superconductivity in Fe based oxy-pnictides i.e., REFeAsO/F has attracted huge attention of the scientific community [2]. In fact, seemingly it was like second revolution after the invention of high $T_c$ cuprates superconductivity way back in 1986 [3] Though other superconductors like famous $MgB_2$ keep on adding [4], but it is only after the Fe pnictides that search for new superconductors is sparked once again. Besides several Fe pnictide [5, 6] and chalcagonide [7, 8] families some other similar structure compounds viz. the $BiS_2$ based [9-11] ones are added more recently. When look-ing for new superconductors, one aspires for best superconducting properties, such as higher critical temperature ($T_c$) and upper critical field ($H_{c2}$). In general, it is a feast for a scientist to discover a higher $T_c$ and $H_{c2}$ superconductor. Both parameters are important for practical applications of a superconductor. Besides the higher $T_c$, which helps in reducing the operating temperature cost, the upper critical field warrants the robustness of the superconductor against magnetic field. In this direction, the recently discovered $Nb_2PdS_5$ superconductor [1, 12], though possess comparatively lower $T_c$ of around 6K only, but the same is quite robust against magnetic field.

The robustness of superconductivity does mean here that superconductivity is least affected by the applied magnetic field. Generally speaking, the decrease of $T_c$ with applied magnetic field i.e. $dT_c/dH$ determines the upper critical field in a simple single band scenario [13]. In a type II superconductor the upper critical field can be enhanced by both intrinsically [6, 14] and extrinsically [15]. Intrinsically, the same happens in multi band systems having complicated electronic [6, 14] and extrinsically by pinning the vortices in their mixed state [15]. The recently discovered superconductor i.e., $Nb_2PdS_5$ seems to be both intrinsically robust as well as pinned by inherent defects etc. It is noteworthy here that Nb is the most popular superconductor being used till date in superconductor industry having varying superconducting transition temperature ($T_c$) values in various metal or alloy forms from 9 K(Nb), 11.5 K(NbC), 16 K(NbN), 19K($Nb_3Sn$) to 23 K($Nb_3Ge$) [16-20].

The recent compound i.e., $Nb_2PdS_5$ somehow reminds not only the importance of Nb being the part of various superconductors in past [16-20], but also the presence of S in its layered structure calls for the attention of famous high upper critical field S containing Chevrel phase compounds [21]. Keeping in view the importance of Nb and the utility of S in new layered structure of the recent compound $Nb_2PdS_5$, we present here an easy and clear route of synthesis of this new superconductor. Further full superconducting characterization in terms of high field (140kOe) and low temperature (down to 2K) electrical, magnetic and thermal characterization are presented in current communication.

**Experiment:**

Polycrystalline bulk $Nb_2PdS_5$ is synthesized via solid state reaction route by quartz vacuum encapsulation technique at $850^0C$. The high purity (4N) ingredients i.e., Nb, Pd and S are weighed in stoichiometric ratio and mixed thoroughly in a glove box in Ar atmosphere. The mixed powder is pelletized, sealed in an evacuated quartz tube and put immediately in box furnace. The sample is heated to $850^0C$ with heating rate of $3^0$ minute and hold at same temperature for 24 hours. Subsequently the furnace is switched off and allowed to cool to room temperature over a span of 6 hours. Thus obtained sample is once again pulverized, pelletized, sealed in encapsulated quartz tube, heated at same temperature for another 24 hours and furnace cooled to room temperature. The room temperature X-ray diffraction (*XRD*) patterns are taken after each heat treatment on Rigaku *XRD* machine. The electrical, magnetic and thermal characterization is done with help of Quantum Design (*QD*) Physical Property Measurement System (*PPMS*) – 140 kOe down to 2K. Hydrostatic pressures are generated by a BeCu/NiCrAl clamped piston-cylinder cell using HPC-33, an addition to PPMS QD, to study the pressure effect on $Nb_2PdS_5$. The pressure at low temperature was calibrated from the superconducting transition temperature of Pb. The sample is immersed in a fluid pressure transmitting medium of Fluorinert in a Teflon cell. Annealed Pt wires were affixed to gold-sputtered contact surfaces on each sample with silver epoxy in a standard four-wire configuration.

**Results and Discussion:**

The synthesized $Nb_2PdS_5$ sample is gray in color. The room temperature fitted and observed X-ray diffraction (*XRD*) pattern for the synthesized $Nb_2PdS_5$ along with its unit cell is shown in Figure 1. $Nb_2PdS_5$ crystallizes in Centro-symmetric structure within space group *C*2/*m* (#12). There are two formula units in each primitive cell comprising two sites each for Nb (Nb1, Nb2) and Pd (Pd1, Pd2) along with five S sites (S1, S2, S3, S4, S5). The schematic sketch of the unit cell is shown in inset of Fig.1. Details of Wyckoff positions, site symmetry and fractional occupancies being obtained from Reitveld fitting of the experimental XRD data are given in Table 1. The lattice parameters are $a = 12.134(2)$Å, $b = 3.277(1)$Å, $c = 15.023(3)$Å, with $\alpha=90°$, $\beta=103.23°(1)$ and $\gamma=90°$. Structural detail results are in general agreement with another recently discovered similar superconducting compound $Ta_2PdS_5$ [22].

In a very recent electronic structure determination theoretical article [23], slightly different configuration is used, i.e., high conductivity direction is labeled along *c*-axis instead of usual *b*-axis [1, 22]. In fact due to anisotropic nature of the conductivity in different Pd-S chains/sheets the superconducting upper critical field is highly directional dependent in these compounds [1, 22, 23]. In a more recent communication [24], the $Nb_2Pd_xS_{5-y}$ fiber exhibited slightly modified superconducting parameters including higher $T_c$ of 7.14K. The role of off stoichiometry and directional alignment could lead to changed superconducting characteristics. Here we stick to stoichiometric $Nb_2PdS_5$ bulk polycrystalline compound to minimize the complications.

*AC* susceptibility versus temperature $\chi(T)$ behavior of the $Nb_2PdS_5$ sample is depicted in Figure 2. *AC* susceptibility measurements are done at 33Hz and 10Oe *AC* drive field. *DC* applied field is kept zero to check the superconducting transition temperature ($T_c$). Both the real ($\chi'$) and imaginary ($\chi''$) part of *ac* susceptibility are measured. Real part ($\chi'$) susceptibility shows transition to diamagnetism at around 5.8K, confirming bulk superconductivity. The imaginary part on the other hand exhibits a single sharp peak in positive susceptibility at around the same temperature. Presence of single sharp peak in $\chi''$ is reminiscent of better superconducting grains coupling in studied $Nb_2PdS$ superconductor. We also measured the *AC* susceptibility of the compound in applied *dc* fields of up to 120kOe. The plot of real part ($\chi'$) of *ac* susceptibility with applied field is depicted in inset of Figure 2. The temperature is fixed at 3K, i.e., well below the superconducting transition temperature of 6K. It is clear from inset of Fig. 2, that the studied $Nb_2PdS_5$ is bulk superconducting (diamagnetic) at 3K in up to 100 kOe applied field. The upper critical field ($H_{c2}$) is above 100kOe just 3K below the superconducting transition temperature ($T_c$). This is first indication that $Nb_2PdS_5$ is quite robust superconductor against magnetic field. More will be discussed after magneto-transport results in next sections.

DC magnetic susceptibility of $Nb_2PdS_5$ sample is shown in Figure 3. The magnetization is done in both *FC* (Field cooled) and *ZFC* (Zero-field-cooled) protocol under applied magnetic field of 10Oe. The compound shows superconducting onset, in terms of FC and ZFC bifurcation, from 5.7K. This is clear from the zoomed inset of Figure 3. There is evidence for substantial flux trapping too. The bifurcation of *FC* and *ZFC* below $T_c$ marks the irreversible region. The shielding fraction as evidenced from *ZFC* diamagnetic susceptibility is quite appreciable (~ 58%). An interesting fact we found repeatedly in case of $Nb_2PdS_5$ superconductor is the appearance of paramagnetic meissner effect (*PME*) in *FC* magnetization. *PME* generally appears in heavily pinned superconductors [25] and could be a good prior indication of high $H_{c2}$.

Figure 4 depicts the resistivity versus temperature ($\rho$-$T$) measurement with and without applied magnetic field. The resistance of the sample decreases with temperature and confirms superconductivity with onset $T_c$ ~ 6.3K and $T_c$ ($\rho = 0$) at 5.9K. The normal state conduction is of metallic type. From the fitting of resistivity $\rho=\rho_o+AT^2$ for low temperatures ($T_c<T<50$ K), we obtained residual resistivity $\rho_o$=1.27 m$\Omega$-cm and A=1.126×10$^{-4}$ m$\Omega$-cm/K$^2$ values. The quadratic variation in resistivity with temperature is shown as red solid curve in Fig.4, which is a clear indication of Fermi-liquid nature of $Nb_2PdS_5$. To study the impact of applied magnetic field on superconductivity of $Nb_2PdS_5$ compound, the magneto-transport measurements are carried out in superconducting region, and the results are shown in inset – I of Fig. 4. With applied field of 110kOe, the $T_c$ ($\rho = 0$) decreases from 5.9K (zero-field) to 2.4K (110KOe). As sketched in inset –II of Fig.4, we have estimated upper critical field $H_{c2}(T)$ by using the conservative procedure of 90%, 50% and 10% dropping of resistivity from normal state into the superconducting transition line. While the applicability of *WHH* (Werthamer-Helfand-Hohenberg) approxi-mation can be debated in this new superconductor, a simplistic single band extrapolation leads to $H_{c2}(0)$ (= -0.69 $T_c$ $dH_{c2}/dT|_{Tc}$) value of 240kOe, 185kOe and 170kOe with of 90%, 50% and 10% criterion respectively. These values match with the experimentally ascertained bulk diamagnetic response of $Nb_2PdS_5$ superconductor at 3K in 100kOe field (inset Fig.2). The estimated values of upper critical field are close to the earlier reported values [12]. Interestingly the $H_{c2}(0)$ value for $Nb_2PdS_5$ superconductor is more than the one as being expected within Pauli Paramagnetic limit of $H_{c2}(0) = 1.84T_c$ of around 110kOe.

The relative decrement of superconductivity with applied field i.e., $dT_c/dH$ is around 0.3K/10 kOe from absolute $T_c$ ($\rho = 0$) criteria. When this is compared with the best of *HTSc* cuprates (3K/10kOe) [26], Fe pnictieds (1K/10kOe) [6, 14], pinned $MgB_2$ (1.5K/10kOe) [15], or the Fe chalcegonides (0.4K/10kOe) [7, 8] and Chevrel phase (0.45 K/10 kOe) [27] compounds, one finds that $Nb_2PdS_5$ is the most robust superconductor against magnetic field till date.

Figure 5 shows the heat capacity ($C_P$) versus temperature ($T$) plots at different applied fields of 0, 10, 50 and 140 kOe in superconducting region i.e., 2-8K for studied $Nb_2PdS_5$ superconductor. In normal state i.e., at 200K the value of $C_P$ is around 180 J/mole K, see inset of Figure 5. The superconducting transition temperature ($T_c$) is seen clearly as a hump in $C_p$ at around 6K. As evident from Fig. 5, the $C_p$ peak temperature shifts to lower temperatures with application of external field and is namely at 5.5K, 5.2K and 4.2K at 0, 10kOe and 50kOe respectively. The $C_p$ hump and associated peak temperature is not seen down to 2K at 140 kOe magnetic field. This shows the compound is not superconducting down to 2K in applied field of 140 kOe.

Figure 6 shows the low temperature normal-state 140KOe $C_p(T)$ data from 2-10K, which is fitted to the Sommerfield–Debye expression as $C_p(T) = \gamma T + \beta T^3 + \delta T^5$. The values of Sommerfield constant ($\gamma$) and $\beta$ are obtained from the fitting of the experimental data are as $\gamma$ = 32mJ/mol-$K^2$, $\beta$ = 1.79mJ/mol-$K^4$ and $\delta$ = -3.48x$10^{-3}$mJ/mol $K^6$. The heat capacity jump ($\Delta C = C_s - C_p$) is estimated by subtracting superconducting heat capacity ($C_s$) to the normal state fitted $C_p$ and the same is shown in the inset of Fig. 6. The normalized value of jump ($\Delta C/\gamma T_c$) is ~ 0.75, which is less than in comparison to the Bardeen–Cooper–Schrieffer (*BCS*) value of 1.43. As seen from inset of Fig. 6, the $C_p$ peak exhibits multiple gap superconductivity signatures with three different $T_c$, with slight upturn near 2K. The relatively higher value of $\gamma$ = 35mJ/mol-$K^2$ is an indication that the $Nb_2PdS_5$ is a strongly coupled superconductor [23]. Worth mentioning is the fact that $\beta$ and $\gamma$ values change slightly, when instead of extended low *T* fitting of experimental $C_p$ to normal state, the 140kOe low *T* experimental data are taken as normal state. Yet, the fact remains that the normalized value of $C_p$ jump is smaller than *BCS* value and the same clearly demonstrates the multi gap superconductivity. The Debye temperature is calculated by using $\Theta_D = (234zR/\beta)^{1/3}$, here $z$ being number of atoms per formula unit and $R$ is the gas constant. Taking the fitted value of $\beta$ (1.79mJ/mol-K$^4$), the calculated value of $\Theta_D$ is 263K.

In order to understand the carries contribution to transport properties in un-doped Nb$_2$PdS$_5$, Hall resistivity ($\rho_{xy}$) and Hall coefficient ($R_H$) measurements were carried out in normal state at different temperatures. Figure 7 depicts the Hall resistivity ($\rho_{xy}$) results for studied Nb$_2$PdS$_5$ superconductor at various temperatures of 250K, 200K, 100K, 50K and 10K up to applied magnetic field of 80kOe. The magnetic field is applied perpendicular to the current flow direction and the voltage is recorded by five probe resistivity measurement method to minimize the magneto resistance contribution. It is clear from Fig. 7 that holes dominate the transport below temperatures 100K, whereas electrons at and above 100K. Clearly this indicates that Nb$_2$PdS$_5$ exhibits a change in type of dominating charge carriers at around 100K. To assert this behavior further the Hall coefficient ($R_H$) of Nb$_2$PdS$_5$ sample is measured under three different applied magnetic fields 5kOe, 30kOe and 50kOe in the temperature range of 2 to 300K, which is shown in Fig. 8. It can be observed from Fig. 8 that the $R_H$ changes its sign from negative (electron type) to positive (hole type) around 190K for all the three fields just after meeting together. $R_H$ is known to be sensitive to curvature of Fermi surface and renormalized Fermi velocity which may lead to change in the sign of Hall coefficient with temperature [28]. Alternatively this behavior can be understood by compensated two band model as proposed by F. Rullier-Albenque et.al to explain transport behavior of LiFeAs [29]. This compensated two band transport property in normal state corroborates multiband superconductivity as claimed by heat capacity study for Nb$_2$PdS$_5$ compound. Spectroscopic investigations need to further furnish this claim. It is also observed that in normal state the maximum of $R_H$ (i.e. carrier concentration, n, is minimum) is at $T_{RH}$(max) ~ 80K even for three different magnetic fields though value of $R_H$ differs. It is interesting to note that this temperature [$T_{RH}$(max)] is very close to temperature at which Hall resistivity ($\rho_{xy}$) changes its sign from negative to positive and not at which $R_H$ changes its sign. Hall coefficient ($R_H$) sharply peaked to maximum at superconducting transition as shown in the inset of Fig.8. For the applied field of 5 kOe, $R_H = 7.01$ cm$^3$/C which is equivalent to hole concentration n = $1/e.R_H = 8.92\times10^{21}$/cm$^3$. Similar type of carrier sign change is seen in another related superconductor 2H-NbS$_2$ [28, 30]. Interestingly, 2H-NbS$_2$ is a known superconductor with $T_c$ above 6K [28, 30, 31]. This prompted us to check if the reported superconductivity of Nb$_2$PdS$_5$ could arise possibly due to minor 2H-NbS$_2$ phase. To exclude the possibility of 2H-NbS$_2$ impurity phase driven superconductivity in Nb$_2$PdS$_5$, we synthesized the NbS$_2$ compound by same heat treatment of quartz vacuum encapsulation and heating of up to 850$^0$C. Thus obtained NbS$_2$ is crystallized in $R3m$ space group with *a* and *c* parameters of 3.335 Å and 17.86 Å [32] and is not superconducting down to 2K, plots not shown. This clearly excluded the possibility of appearance of $NbS_2$ impurity driven superconductivity in $Nb_2PdS_5$. To further the investigation of transport properties of $Nb_2PdS_5$ resistivity as a function of temperature, $\rho(T)$, is measured under various hydrostatic pressures up to 1.98 GPa which is shown in Fig.9. There is no appreciable change on $T_c$ as can be seen from the inset of Fig. 9. The normal state resistivity is though slightly increased for lower pressures of up to 0.97GPa, the same exhibits decreasing trend for 1.68 and 1.97GPa pressures. On the other hand we found nearly no change in $T_c$.

**Conclusions:**

In conclusion we have synthesized the new layered sulfide $Nb_2PdS_5$ superconductor and established its bulk superconductivity by magnetization, transport and heat capacity measurements below 6K, with high upper critical field outside Pauli paramagnetic limit. The $C_p$ measurements exhibited multiple gap superconductivity signatures with three different $T_c$. The Sommerfeld constant ($\gamma$) and Debye temperature ($\Theta_D$) as determined from low temperature fitting of $C_P(T)$ are 35mJ/mole-$K^2$ and 253K respectively. Hall studies in conjugation to heat capacity measurements affirm the possibility of multiband superconductivity in $Nb_2PdS_5$.

**Acknowledgement:**

Authors would like to thank their Director Prof. R.C. Budhani for his keen interest in the present work. This work is supported by DAE-SRC outstanding investigator award scheme to work on search for new superconductors. H. Kishan thanks CSIR for providing Emeritus Scientist Fellowship.

**Figure captions:**

Figure. 1 (color online) Rietveld refined Room temperature $X$-ray diffraction ($XR$D) patterns of $Nb_2PdS_5$, the unit cell of the compound is shown in inset.

Figure. 2 (color online) $AC$ susceptibility $\chi(T)$ behavior of the $Nb_2PdS_5$ sample at frequency 333Hz and $ac$ drive amplitude of 10Oe, the inset shows the real part of the same in applied dc fields of up to 120kOe

Figure. 3 (color online) Temperature variation of $DC$ Magnetization in $ZFC$ and $FC$ modes for $Nb_2PdS_5$ compound at 10Oe, inset shows the expanded part of the same plot indicating irreversible behavior and marking of $T_c$.

Figure. 4 (color online) Resistivity vs. temperature ($\rho$-$T$) behavior of $Nb_2PdS_5$, inset – I shows the same in various applied fields of 0 – 110kOe below 10K and inset – II is the upper critical field estimated from the $\rho(T)H$ data with 90%, 50% and 10% $\rho_n$ criteria.

Figure. 5 (color online) Heat capacity ($C_p$) versus temperature plot for $Nb_2PdS_5$ superconductor; inset shows the same in 0, 10, 50 and 140kOe fields in superconductivity region below 8 K.

Figure. 6 (color online) Fitted (red line) and observed (filled squares) $Cp$ versus $T$ data for $Nb_2PdS_5$ at applied field of 140kOe in low temperature region, the inset shows the electronic specific heat peak at $T_c$.

Figure. 7 Hall resistivity ($\rho_{xy}$) versus magnetic field plots at various temperatures of 250K, 200K, 100K, 50K and 10K for $Nb_2PdS_5$ superconductor.

Figure. 8 Hall coefficient ($R_H$) versus temperature plot for $Nb_2PdS_5$ superconductor under applied fields of 5kOe, 30kOe and 50kOe. 5kOe $R_H$ ($T$) is multiplied by 1/5 for the clarity of viewgraph. Magnified graph close to superconducting transition is shown in inset.

Figure. 9 Resistivity under different hydrostatic pressure (0.35-1.98GPa) as function of temperature for $Nb_2PdS_5$ superconductor, inset shows the magnified view of the same close to superconducting transition.

**Table 1** Reitveld refined Wyckoff positions and fractional occupancies of the atoms in $Nb_2PdS_5$

| Atom | x | y | z | site | Fractional occupancy |
|------|---|---|---|------|----------------------|
| Nb1 | 0.0759(5) | 0.5000 | 0.179(4) | 4i | 1 |
| Nb2 | 0.1528(8) | 0.0000 | 0.3783(2) | 4i | 1 |
| Pd1 | 0.0000 | 0.0000 | 0.0000 | 2a | 1/2 |
| Pd2 | 0.0000 | 0.0000 | 0.5000) | 2c | 1/2 |
| S1 | 0.3503(6) | 0.0000 | 0.4890(1) | 4i | 1 |
| S2 | 0.2537(4) | 0.5000 | 0.2950(6) | 4i | 1 |
| S3 | 0.1753(6) | 0.0000 | 0.0977(8)) | 4i | 1 |
| S4 | 0.4230(3) | 0.5000 | 0.1321(4) | 4i | 1 |
| S5 | 0.4904(7) | 0.0000 | 0.3234(6) | 4i | 1 |

Fig. 1

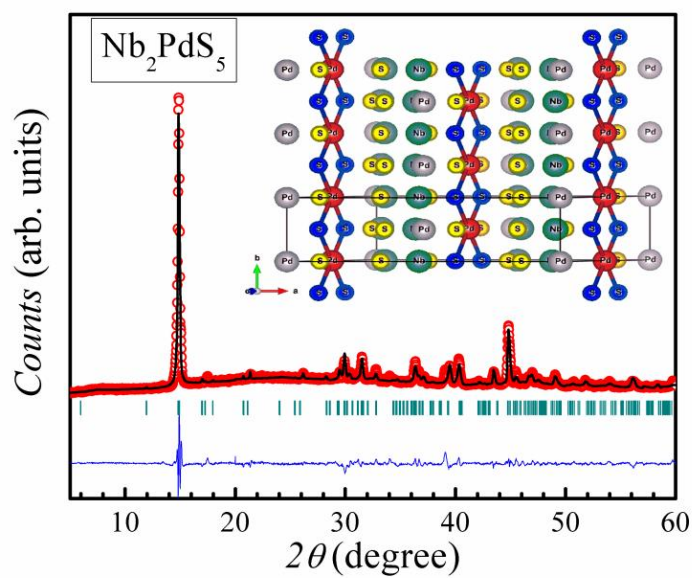

Fig. 2

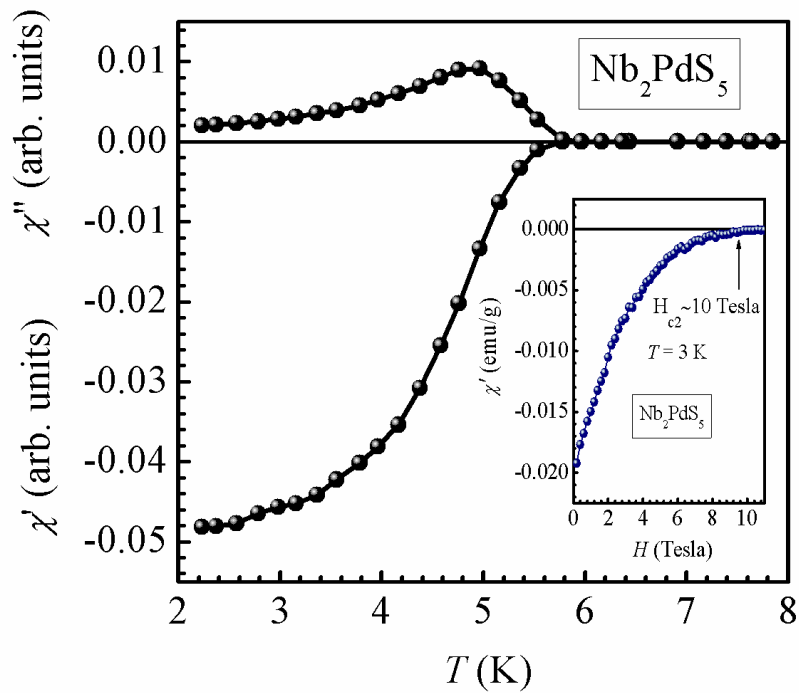

Fig. 3

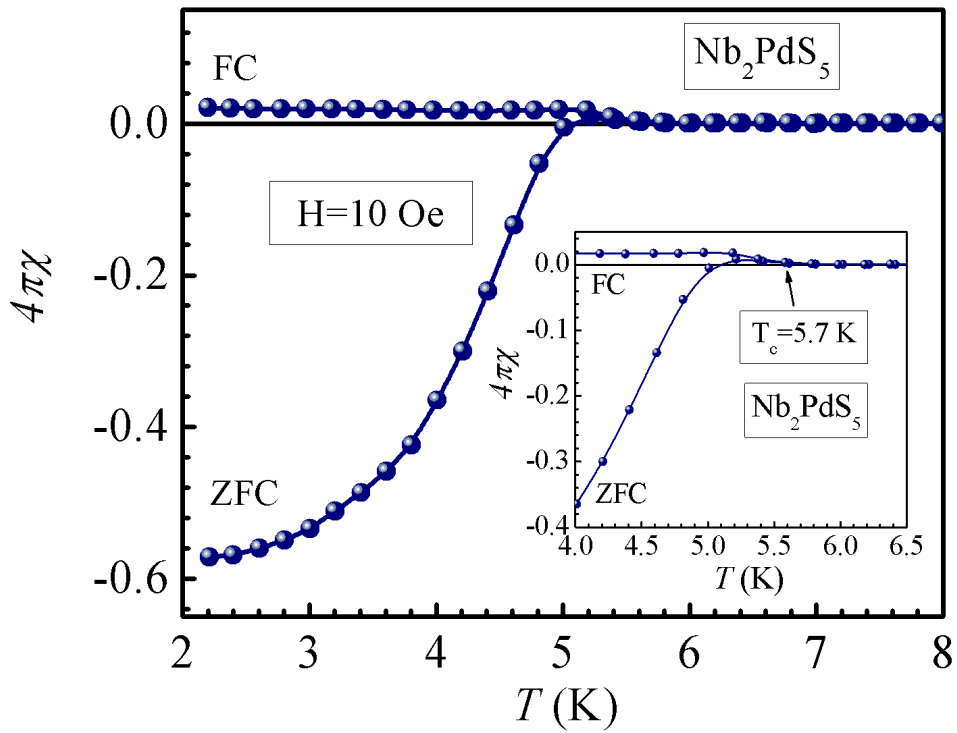

Fig. 4

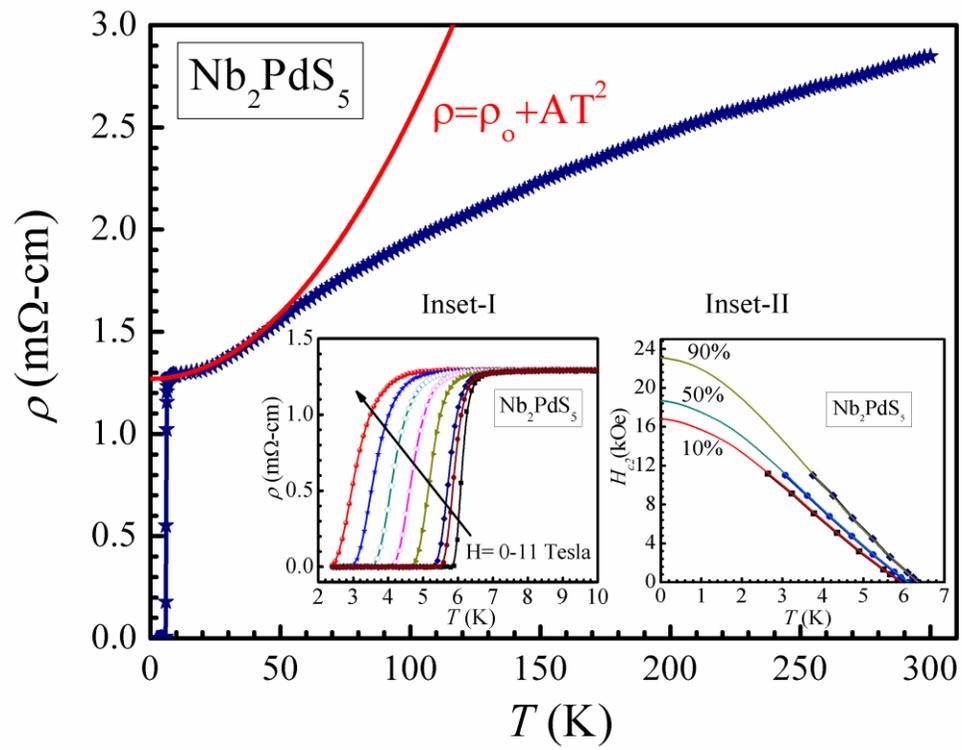

Fig. 5

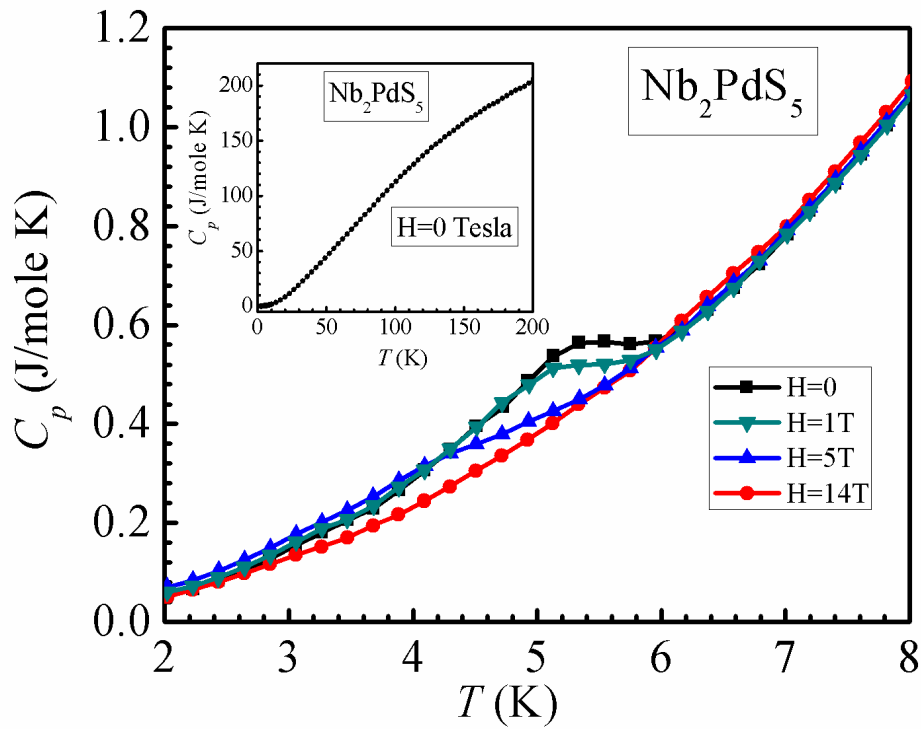

Fig. 6

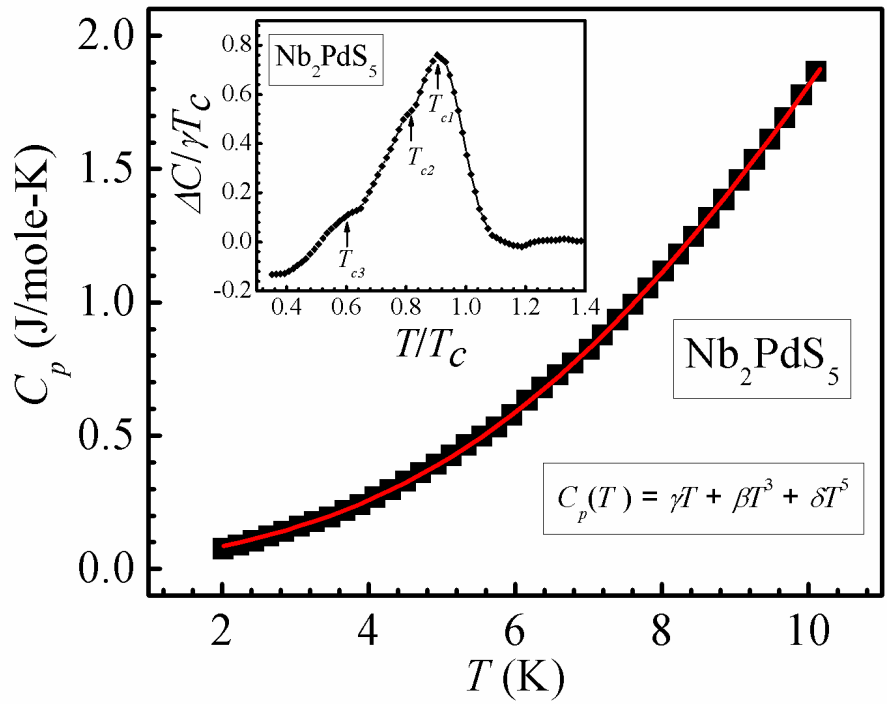

Fig. 7

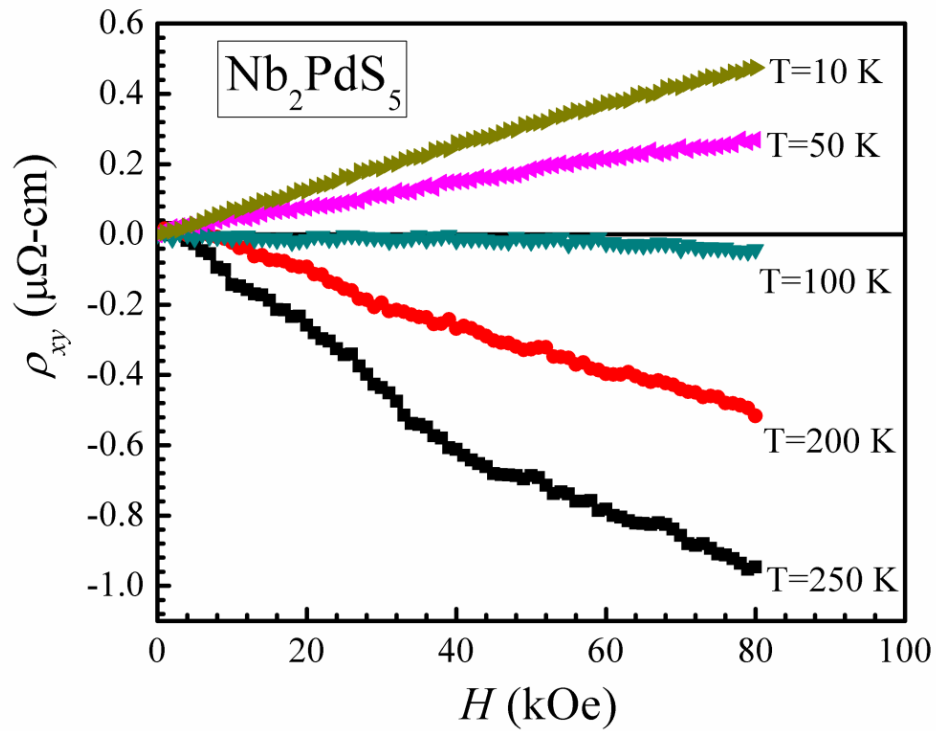

Fig. 8

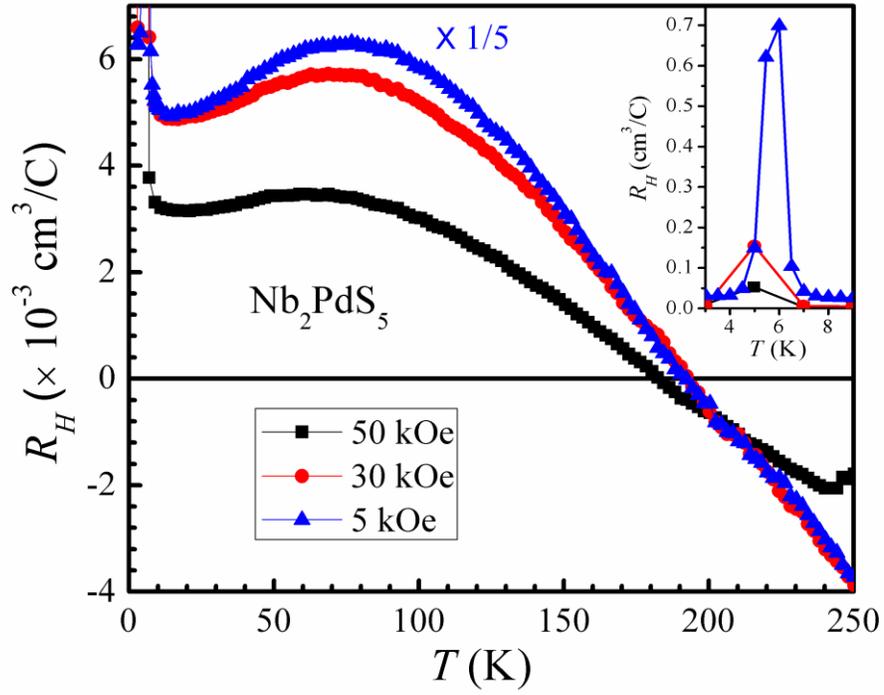

Fig. 9

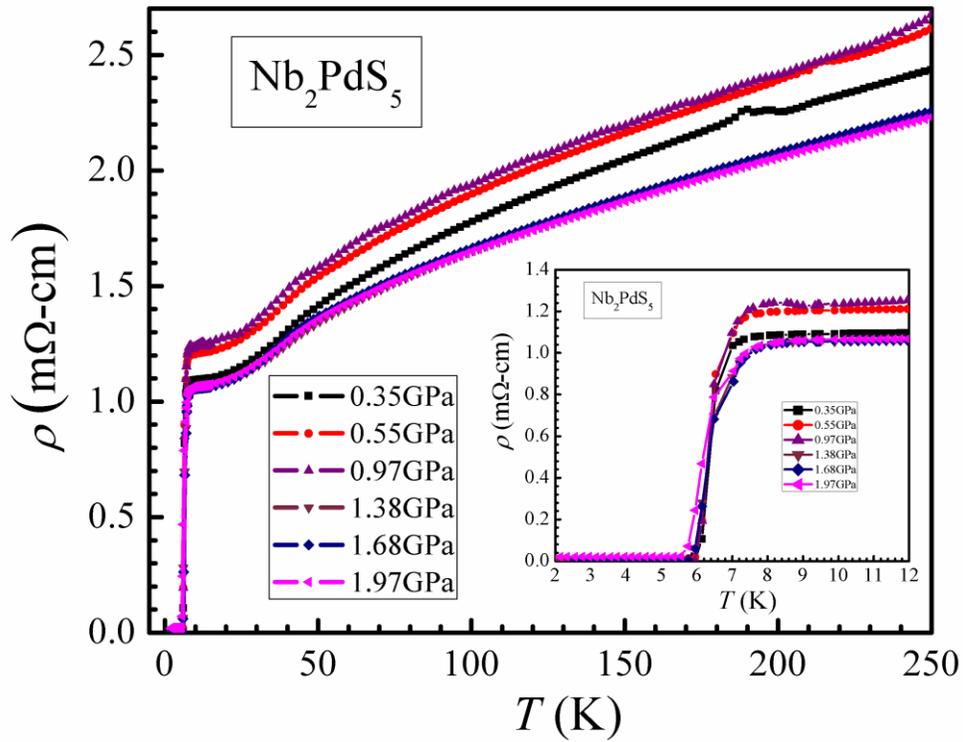